\documentclass{iau}
\usepackage[pdftex]{graphicx}
\usepackage{floatrow}
\usepackage{wrapfig}

\title[AGNfitter: Fitting SED of AGN] %% give here short title %%
{Fitting Spectral Energy Distributions\\ of AGN\\ A Markov Chain Monte Carlo Approach}

\author[Calistro Rivera et al.]   %% give here short author list %%
{Gabriela Calistro Rivera$^1$, Elisabeta Lusso$^1$, Joseph F. Hennawi$^1$, David W. Hogg$^2$}

\affiliation{$^1$Max Planck Institute for Astronomy, K\"onigstuhl 17, 69117, Heidelberg, Germany %\\ email: {\tt m.lugaro@phys.uu.nl} 
\\[\affilskip]
$^2$Center for Cosmology \& Particle Physics, Department of Physics, New York University, USA} % \\email: {\tt hoefner@astro.uu.se}}

\pubyear{2013}
\volume{134}  %% insert here IAU Symposium No.
\pagerange{119--126}
\jname{Multiwavelength AGN Surveys and Studies}
\editors{A.C. Editor, B.D. Editor \& C.E. Editor, eds.}

\begin{document}

\maketitle
\begin{abstract}
We present AGN\textit{fitter}: a Markov Chain Monte Carlo algorithm developed to fit the spectral energy distributions (SEDs) of active galactic nuclei (AGN) with different physical models of AGN components. This code is well suited to determine in a robust way multiple parameters and their uncertainties, which quantify the physical processes responsible for the panchromatic nature of active galaxies and quasars. We describe the technicalities of the code and test its capabilities in the context of X-ray selected obscured AGN using multiwavelength data from the XMM-COSMOS survey.
\vspace{-0.2cm}
%% add here a maximum of 10 keywords, to be taken form the file <Keywords.txt>
\end{abstract}

\firstsection % if your document starts with a section,
\firstsection
%\firstsection           
    % remove some space above using this command.

\section{Motivation}
The radiation emitted by physical processes ongoing in active galaxies sculpt a spectral energy distribution (SED) that spreads on a wide range of the electromagnetic spectrum. Multi-wavelength photometry is thus a rich source of information about the AGN nature.
%This radiation can be produced at different physical scales in the AGN morphology and it is important to disentangle its origins, in order to infer right physical interpretations from the observations.
One method to extract this information is through SED fitting, which consist in comparing the observed photometric data to a combination of physical models for AGN components. It is customary to perform SED fitting using optimization methods as $\chi^{2}$-minimization, which defines the best fit as the combination of parameters which models the whole SED showing the minimal $\chi^{2}$ value. However, this method is statistically correct only under the assumption that the parameters are fully independent from each other and thus have a Gaussian probability distribution. This assumption is a drawback of this method since the parameters describing AGN physics are in most cases highly degenerated. To solve this issue we present AGN\textit{fitter}, a bayesian SED fitting code for AGN that allows an integral calculation of the posterior probability distributions of the model parameters taking into account degeneracies and correlations existing among them.	
\vspace{-0.6cm}  
\section{Sampling the parameter space}
%\vspace{-0.03cm}
AGN\textit{fitter} samples the parameter space built by the AGN models' parameters using a Markov Chain Monte Carlo method. This consists in a random walk that is biased for regions of higher probability in the parameter space, making the code fast and efficient since no time is lost in non-interesting regions. Built on the published code Emcee (\cite[Foreman-Mackey et al. 2011]{D. Foreman Mackey}) our MCMC code increases its efficiency taking advantage of multiprocessing by parallel tempering, i.e. exploring the parameter space with several chains simultaneously.
\begin{figure}[h!]%{width=0.5\textwidth}
\hspace{-0.35cm}
\vspace*{-0.37 cm}
\floatbox[{\capbeside\thisfloatsetup{capbesideposition={left,center},capbesidewidth=4.3cm}}]{figure}[0.96\FBwidth]{\caption{Eight different realizations of the complete SED of one source are drawn as red lines. These SEDs are constructed from eight different combinations of parameters, i.e. different points in the multiparameter space. A decomposed version of each SED is drawn in other colors. The green and yellow lines represent the cold dust and galaxy radiation components, while the purple and blue represent the AGN radiation emmited by the hot dust region and the accretion disk respectively.}\label{fig:test}}
{\includegraphics[width=9.0cm]{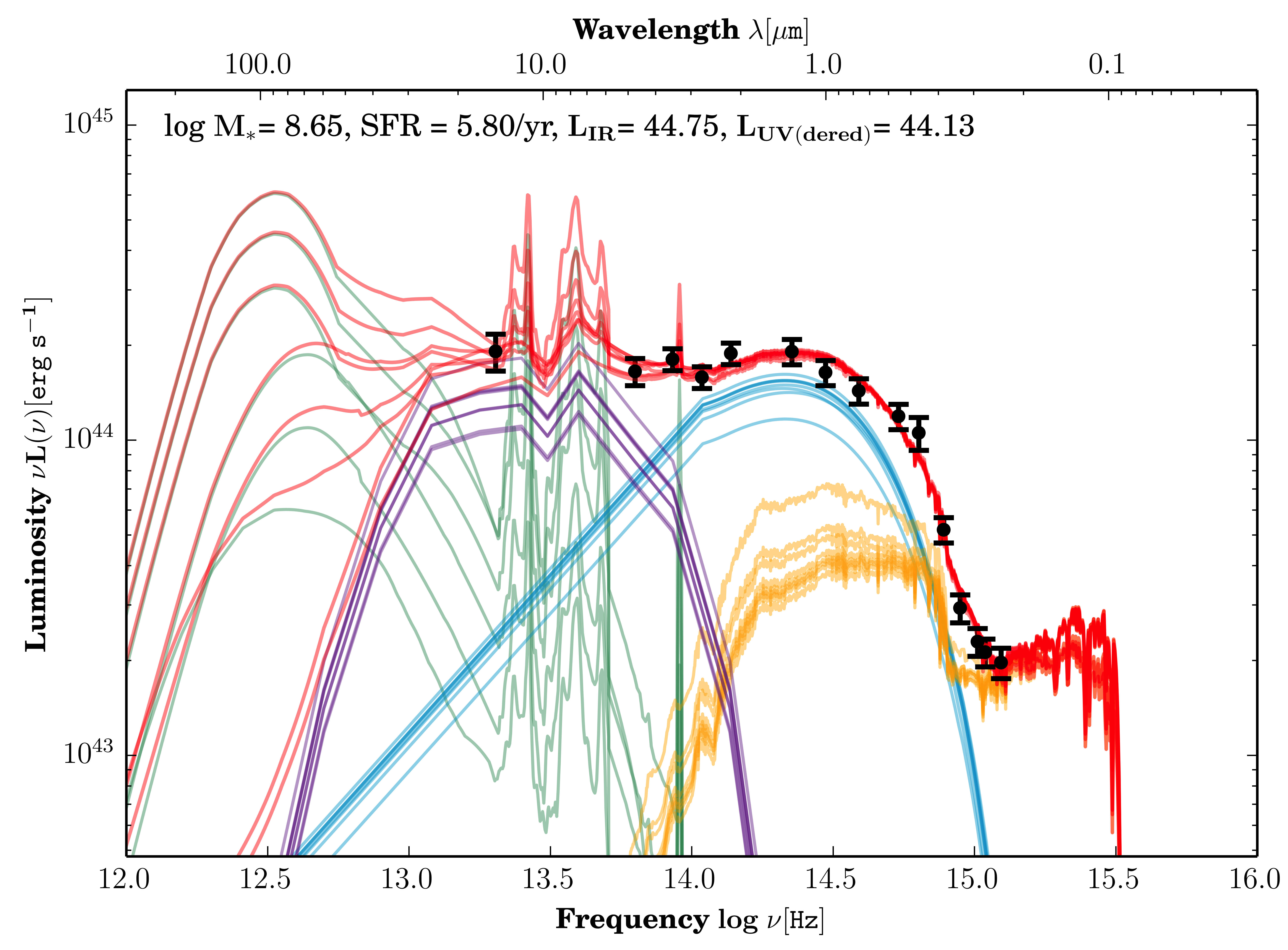}}
%\begin{center}
%\includegraphics[width=0.8\textwidth]{SED_242} 
%\vspace*{-1.0 cm}
%\caption{fsdfds}
%\label{fig2}
%\end{center}
%\hspace*{-3 cm}
\end{figure}
The dimension of the parameter space sampled by AGN\textit{fitter} is constructed in this first version by 10 parameters, which rule the modeling of four AGN components: the accretion disk radiation, the nuclear hot dust emission and the radiation emitted by the host galaxy and the star burst regions. For the accretion disk radiation (big blue bump) we use the model by \cite[Richards et al. (2006)]{Gordon Richards}, while the hot dust surrounding the disk is modeled by a continuous torus (\cite[Silva et al. 1994]{Silva}). The contribution of the host galaxy to the full source radiation is modeled using \cite[Bruzual \& Charlot (2003)]{Bruzual and Charlot} templates, while the cold dust radiation produced in star burst regions is simulated using \cite[Dale \& Helou (2002)]{Dale} templates. Since many of these models cover equal regions of the spectrum it is important to take correlations into account. %these parameters are not completely uncorrelated and it is important to take this into account when calculating uncertainties of the resulting models.
\vspace{-0.6cm}  
\section{The results of AGN\textit{fitter}}
\begin{wrapfigure}{l}{0.48\textwidth}
%[h]
\vspace{-0.5cm}
\includegraphics[width=1\textwidth]{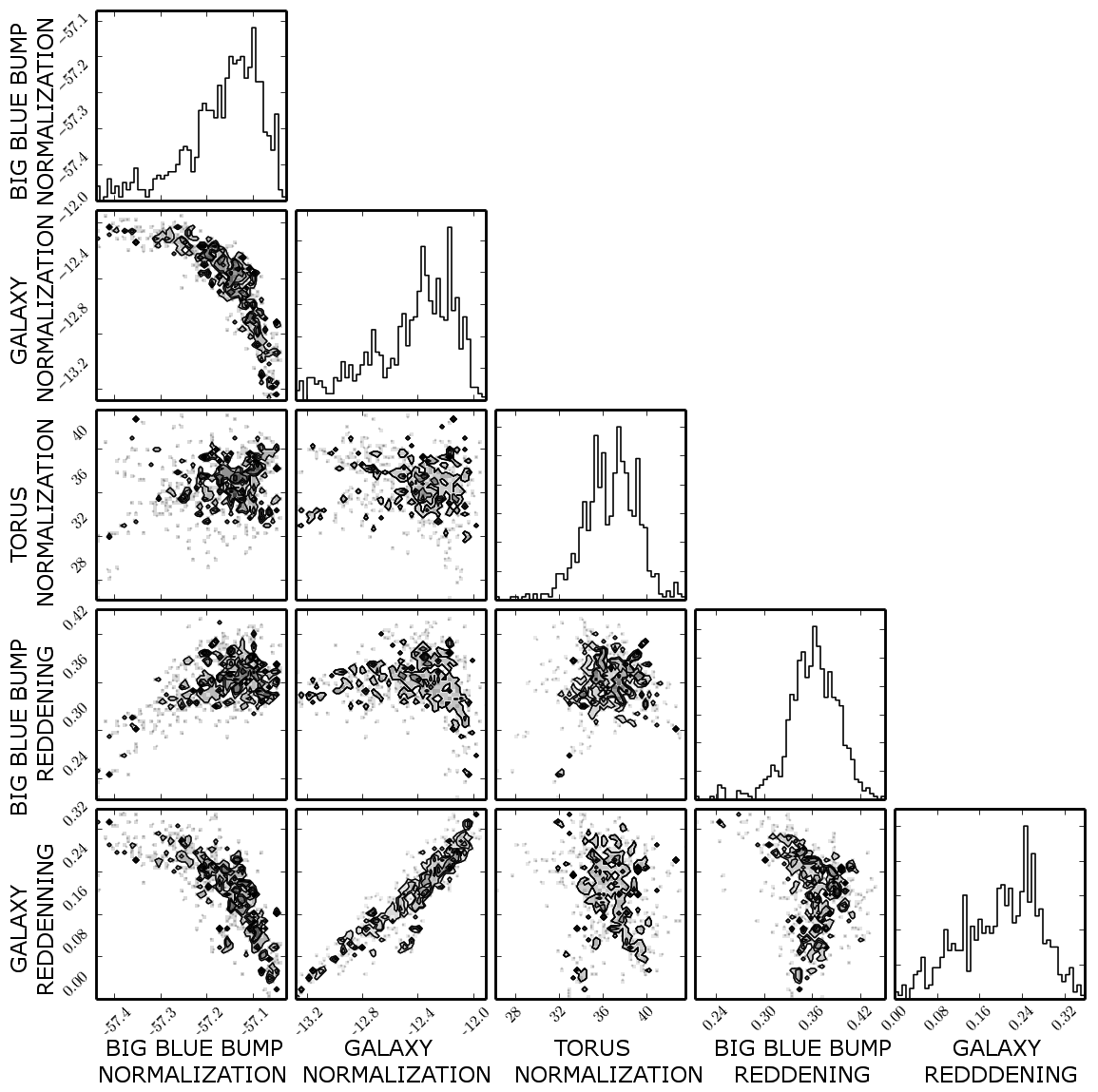}
%\hspace*{1cm}
\vspace*{-0.7 cm}
\caption{Marginalized and joint posterior density functions (PDFs) of five physical parameters constrained from the SED fitting process. Degeneracies between AGN and galaxy components can be clearly recognized.}%(big blue bump) and galaxy components can be recognized.}
\label{fig1}
%\end{center}
\end{wrapfigure}
%AGN\textit{fitter} uses photometrical data as input, explores a ten dimensional parameter space in order to find fitting models for the data and provides a full information of the posterior probability of each model, revealing many SED models which similar fitting probability. 
AGN\textit{fitter} constructs SEDs from existing physical models (Fig. 1) and calculate physical parameters that are interesting to AGN physics, such as relevant integrated luminosities ($L_{bol}$,  $L_{FIR}$, $L_{IR}$) and parameters ruling the physics of the host galaxy, such as age, stellar mass $M_*$ and star formation rate (SFR).
Moreover, AGN\textit{fitter} provides both the marginalized and two-dimensional posterior density functions (PDF) of the parameters listed above. In this way degeneracies can not only be better visualized but also analyzed and treated (Fig.2). %Fig. shows an example for the marginalized and two-dimensional PDF for the parameters ruling the age and star formation history in the galaxy model and the parameter ruling the luminosity of the big blue bump. One can recognize the anti-correlation between the values of these parameters. 
Finally, due to the code's Bayesian methodology, the user is able to take advantage of prior constraints on the parameters' distributions. In this way the information given by the likelihood function can be complemented, calculating robustly posterior probabilities of the parameters. 
%\textit{Testing the results: }In order to test the robustness of our procedure we have run AGN\textit{fitter} for a sample of AGN (both obscured and unobscured) from the XMM-COSMOS survey. and used the same photometric data to model the SEDs using a different method for SED-fitting developed by \cite[Lusso et al. 2010]{Beta}. To ensure a reasonable comparison we use the same number of parameters and models in both methods, we restrict AGN\textit{fitter} to flat priors and calculate likelihood functions instead of posterior probabilities. Both methods obtain very similar values of the parameters of interest, the differences being found as expected by sources, which show high degrees of degeneracies.
\vspace{-0.6cm} 
\section{Conclusions}
AGN\textit{fitter} is a robust statistical tool to model AGN SEDs and to infer physical parameters from multiwavelength photometrical data. Our code provides a vast statistical information about the inferred parameters allowing in this way an analysis of multiwavelength photometry cognizant of degeneracies and correlations, which is necessary for AGN studies of general purpose.

\end{document}